\documentclass[prd,twocolumn,noshowpacs,superscriptaddress,nofootinbib]{revtex4-1}
\usepackage[T1]{fontenc} % if needed
\usepackage{amsmath}
\usepackage{amssymb}
\usepackage{amsfonts}
\usepackage{graphicx}
\usepackage{enumitem}
\usepackage{bm}
\usepackage{rotating}
\usepackage{hyperref}
\usepackage{xcolor}
\usepackage{array}
\usepackage{lmodern}
\usepackage[normalem]{ulem}
\usepackage{braket}
\usepackage{tensor}
\usepackage{mathrsfs}
\usepackage{float}

\usepackage[normalem]{ulem}

\begin{document}

\title{Minimal Dirac seesaw accompanied by Dirac fermionic dark matter}

\author{Pei-Hong Gu}

\email{phgu@seu.edu.cn}

\affiliation{School of Physics, Jiulonghu Campus, Southeast University, Nanjing 211189, China}

%\affiliation{School of Physics and Astronomy, Shanghai Jiao Tong University, 800 Dongchuan Road, Shanghai 200240, China}

\begin{abstract}

The $SU(3)_c^{}\times SU(2)_L^{} \times U(1)_Y^{}$ standard model is extended by a $U(1)_{B-L}^{}$ gauge symmetry with four right-handed neutrinos. Because of their Yukawa couplings to a Higgs singlet for spontaneously breaking the $U(1)_{B-L}^{}$ symmetry, two right-handed neutrinos can form a Dirac fermion to become a stable dark matter particle. Meanwhile, mediated by additionally heavy Higgs doublet(s), fermion singlet(s) and/or fermion doublet(s), the other two right-handed neutrinos can have a dimension-5 operator with the standard model lepton and Higgs doublets as well as the $U(1)_{B-L}^{}$ Higgs singlet. This context can realize a minimal Dirac neutrino mass matrix only with two nonzero eigenvalues. In association with the sphaleron processes, the interactions for generating the Dirac neutrino masses can also produce the observed baryon asymmetry in the universe.

\end{abstract}

%\pacs{98.80.Cq, 14.60.Pq, 95.35.+d, 12.60.Cn, 12.60.Fr}

\maketitle

\section{Introduction}

The discovery of neutrino oscillations indicates that three flavors of neutrinos should be massive and mixed \cite{navas2024}. Meanwhile, the cosmological observation requires that the neutrinos should be extremely light \cite{navas2024}. The tiny neutrino masses can be naturally induced in various seesaw \cite{minkowski1977} extensions of the $SU(3)_c^{}\times SU(2)_L^{}\times U(1)^{}_{Y}$ standard model (SM). In these popular seesaw scenarios \cite{minkowski1977,mw1980,flhj1989,tao1996,ma2006}, the neutrino mass generation is accompanied by certain lepton-number-violating interactions and hence the neutrinos have a Majorana nature. Meanwhile, the interactions for realizing the seesaw can produce a lepton asymmetry stored in the SM leptons and then the produced lepton asymmetry can be partially converted to a baryon symmetry by the sphaleron processes \cite{krs1985}. This is the so-called leptogenesis mechanism \cite{fy1986} to explain the observed baryon asymmetry in the universe \cite{ma2006,fy1986,lpy1986,fps1995,fpsw1996,crv1996,pilaftsis1997,ms1998,bcst1999,hambye2001,di2002,gnrrs2003,hs2004,bbp2005}.

However, we should keep in mind that the theoretical assumption of the lepton number violation and then the Majorana neutrinos has not been confirmed by any experiments yet. So, it is worth studying the possibility of Dirac neutrinos \cite{rw1983,dlrw1999,mp2002,tt2006,gh2006,gu2016,gs2007,gu2012,gu2017,gu2019-1,es2015,bd2016,aas2017,wh2017,yd2017,csv2018,cryz2019,bdhps2019,saad2019,ekss2019,jks2019,ma2019}. In analogy to the usual seesaw models for the Majorana neutrino mass generation, we can construct some Dirac seesaw models \cite{rw1983,mp2002,gh2006,gu2016,gs2007} for the Dirac neutrino mass generation. The interactions for the Dirac seesaw can induce a lepton asymmetry stored in the SM left-handed leptons and an opposite lepton asymmetry stored in the right-handed neutrinos although the total lepton asymmetry is exactly zero. The right-handed neutrinos will go into equilibrium with the left-handed neutrinos at a very low temperature, where the sphalerons have already stopped working. Therefore, the sphalerons will never affect the right-handed neutrino asymmetry, but it can still transfer the SM lepton asymmetry. This type of leptogenesis is named as the neutrinogenesis mechanism \cite{dlrw1999} and has been studied in literatures \cite{mp2002,dlrw1999,tt2006,gh2006,gu2016,gs2007,gu2012,gu2017,gu2019-1}. In the Dirac seesaw models, the renormalizable Yukawa couplings of the right-handed neutrinos to the SM lepton and Higgs doublets can appear until an additionally discrete, global or gauge symmetry is spontaneously broken. This new symmetry breaking scale may be constrained by other new physics. For example, in a class of mirror models, the additional symmetry is a mirror electroweak symmetry so that it can be fixed by the dark matter mass \cite{gu2012}.

In this paper we shall realize the Dirac seesaw by introducing a $U(1)_{B-L}^{}$ gauge symmetry with four right-handed neutrinos. Because of their Yukawa couplings to a Higgs singlet for spontaneously breaking the $U(1)_{B-L}^{}$ symmetry, two right-handed neutrinos can form a Dirac fermion and then become a stable dark matter particle.  Furthermore, additionally heavy Higgs doublet(s) \cite{gh2006} , fermion singlet(s) \cite{rw1983} and/or fermion doublet(s) \cite{gu2016} can mediate a dimension-5 operator among the other two right-handed neutrinos, the SM lepton and Higgs doublets as well as the $U(1)_{B-L}^{}$ Higgs singlet. This means a highly suppressed Dirac neutrino mass matrix only with two nonzero eigenvalues. Finally, through the interactions for the neutrino mass generation, the heavy Higgs doublet(s), fermion singlet(s) and/or fermion doublet(s) can decay to realize a leptogenesis mechanism.

\section{The model}

The SM fermions and scalar are denoted as follows, 
\begin{eqnarray}
\label{sm}
&&\begin{array}{l}q^{}_{Li}(3,2,+\frac{1}{6})(+\frac{1}{3})\,,\end{array} ~~\begin{array}{l}d^{}_{Ri}(3,1,-\frac{1}{3})(+\frac{1}{3})\,,\end{array}  \nonumber\\
[2mm]
 &&\begin{array}{l}u^{}_{Ri}(3,1,+\frac{2}{3})(+\frac{1}{3})\,,\end{array} ~~\begin{array}{l}l^{}_{L\alpha}(1,2,-\frac{1}{2})(-1)\,,\end{array} \nonumber\\
[2mm]
&&\begin{array}{l}e^{}_{R\alpha}(1,1,-1)(-1)\,,\end{array} ~~\begin{array}{l}\phi^{}(1,2,-\frac{1}{2})(0)\,,\end{array}\nonumber\\
[2mm]
&&(i=1,2,3;~\alpha=e,\mu,\tau.)\end{eqnarray}
Here and thereafter the first and second brackets following the fields respectively describe the transformations under the SM $SU(3)_c^{} \times SU(2)^{}_{L}\times U(1)_Y^{}$ gauge groups and the $U(1)_{B-L}^{}$ gauge group.

In order to cancel the gauge anomalies, we need some right-handed neutrinos \cite{mp2007,ms2015,pry2016}. In the present work, we consider the following right-handed neutrinos,
\begin{eqnarray}
\label{rhn}
&&\begin{array}{l}\nu^{}_{R1}(1,1,0)(-2)\,,\end{array}~~\begin{array}{l}\nu^{}_{R2}(1,1,0)(-2)\,,\end{array}\nonumber\\
[2mm]
&&\begin{array}{l}\nu^{}_{R3}(1,1,0)(\frac{1+\sqrt{17}}{2})\,,\end{array}~~\begin{array}{l}\nu^{}_{R4}(1,1,0)(\frac{1-\sqrt{17}}{2})\,.\end{array}
\end{eqnarray}
Frankly speaking, the irrational $U(1)_{B-L}^{}$ charges are unconventional. However, such theoretical arrangement is not ruled out by any experiments. 
For spontaneously breaking the $U(1)_{B-L}^{}$ symmetry, we introduce a Higgs singlet as below,
\begin{eqnarray}
&&\begin{array}{l}\xi(1,1,0)(+1)\,.\end{array}
\end{eqnarray}
The mass of the $U(1)_{B-L}^{}$ gauge boson $Z_{B-L}^{}$ then should be 
\begin{eqnarray}
M_{Z_{B-L}^{}}^{}&=& \sqrt{2} g_{B-L}^{} \langle \xi\rangle \,,
\end{eqnarray}
with $g_{B-L}^{}$ being the $U(1)_{B-L}^{}$ gauge coupling. The experimental constraints on the $U(1)_{B-L}^{}$ gauge symmetry is \cite{afpr2017,klq2016},
\begin{eqnarray}
\label{low}
\frac{M_{Z_{B-L}^{}}^{}}{g_{B-L}^{} }  \gtrsim 7 \,\textrm{TeV} \Rightarrow \langle \xi \rangle \gtrsim 5\,\textrm{TeV}\,.
\end{eqnarray}

It is easy to see the Higgs singlet $\xi$ can have a Yukawa interaction with the third and forth right-handed neutrinos $\nu_{R3}^{}$ and $\nu_{R4}^{}$, i.e.
\begin{eqnarray}
\label{dm}
\mathcal{L}\supset - y_{34}^{} \left(\xi \bar{\nu}_{R3}^{} \nu_{R4}^{c} +\textrm{H.c.}\right)\,.
\end{eqnarray}
Furthermore, in association with the Higgs singlet $\xi$, we can construct the following dimension-5 operators involving the first and second right-handed neutrinos $\nu_{R1,2}^{}$, i.e.
\begin{eqnarray}
\label{nmass}
\mathcal{L}\supset - \sum_{i=1,2 \atop \alpha=e,\mu,\tau}^{}\frac{c_{ \alpha i}^{}}{\Lambda} \bar{l}_{L\alpha}^{} \phi \nu_{Ri}^{}\xi +\textrm{H.c.}\,.
\end{eqnarray}
As shown later, the above effective operators can be induced only by three types of renormalizable models as follows, 
\begin{eqnarray}
\label{lag}
\mathcal{L}_{\textrm{II}}^{}&\supset&- \sum_{i=1,2 \atop \alpha=e,\mu,\tau}^{a=1,...}\left[M_{\eta_a}^2 \eta^\dagger_{a}\eta_{a}^{} + \rho_{a}^{} \left(\xi \eta^\dagger_{a}\phi +\textrm{H.c.}\right) \right. \nonumber\\
[1mm]
&&\left.+ f_{ a\alpha  i}^{} \bar{l}_{L\alpha}^{} \eta_a^{} \nu_{Ri}^{}+ \textrm{H.c.}\right]\,,\\
[2mm]
\mathcal{L}_{\textrm{I}}^{}&\supset&-\sum_{i=1,2 \atop \alpha=e,\mu,\tau}^{b=1,...}\left[ M_{S_b}^{} \left(\bar{S}_{Rb}^{} S'^{c}_{Rb}+\textrm{H.c.}\right) \right. \nonumber\\
[1mm]
&&\left.+ y_{L\alpha b}^{} \bar{l}_{L\alpha}^{} \phi S_{Rb}^{}+ y_{R bi}^{} \bar{S}'^{c}_{Rb} \nu_{Ri}^{} \xi  + \textrm{H.c.}\right]\nonumber\\
[1mm]
&=& -\sum_{i=1,2 \atop \alpha=e,\mu,\tau}^{b=1,...}\left[ M_{S_b}^{} \bar{S}_{b}^{} S_{b}^{} + y_{L\alpha b}^{} \bar{l}_{L\alpha}^{} \phi S_{b}^{}  \right. \nonumber\\
[1mm]
&&\left.+  y_{R ci}^{} \bar{S}_{b}^{} \nu_{Ri}^{} \xi+ \textrm{H.c.}\right]~~\textrm{with}\nonumber\\
[2mm]
&&S_b^{}=S'^{c}_{Rb}+S^{}_{Rb}\,,\\
[2mm]
\mathcal{L}_{\textrm{III}}^{}&\supset&-\sum_{i=1,2 \atop \alpha=e,\mu,\tau}^{c=1,...}\left[ M_{D_c}^{} \left(\bar{D}'^{c}_{Lc} i\tau_2^{} D_{Lc}^{} + \textrm{H.c.}\right) \right.\nonumber\\
[1mm]
&&\left.+ g_{L\alpha c}^{} \bar{l}_{L\alpha}^{} i\tau_2^{} D'^{c}_{Lc}  \xi + g_{R c i}^{} \bar{D}_{Lc}^{} \phi \nu_{Ri}^{}  + \textrm{H.c.}\right]\nonumber\\
[1mm]
&=&-\sum_{i=1,2 \atop \alpha=e,\mu,\tau}^{c=1,...}\left[ M_{D_c}^{} \bar{D}^{}_{c} D_{c}^{}  + g_{L\alpha c}^{} \bar{l}_{L\alpha}^{} D^{}_{c}  \xi \right.   \nonumber\\
[1mm]
&&\left.+ g_{R c i}^{} \bar{D}_{c}^{} \phi \nu_{Ri}^{}+ \textrm{H.c.}\right]~~\textrm{with}   \nonumber\\
[2mm]
&&D_c^{} = D_{Lc}^{} + i\tau_2^{} D'^{c}_{Lc}\,,
\end{eqnarray}
where $\eta$, $S$ and $D$ are additionally heavy Higgs doublet(s), fermion singlet(s), and fermion doublet(s), 
\begin{eqnarray}
&&\begin{array}{l}\eta(1,2,-\frac{1}{2})(+1)\,;\end{array}\nonumber\\
[2mm]
&&\begin{array}{l}S_{R}^{}(1,1,0)(-1)\,,\end{array}~~ \begin{array}{l}S'^{}_{R}(1,1,0)(+1)\,;\end{array}\nonumber\\
[2mm]
&& \begin{array}{l}D_{L}^{}(1,2-\frac{1}{2})(-2)\,,\end{array}~~\begin{array}{l}D'^{}_{L}(1,2,+\frac{1}{2})(+2)\,.\end{array}\end{eqnarray}
For convenience and without loss of generality, we have chosen the mass matrices $M_\eta^2$, $M_S^{}$ and $M_D^{}$ to be real and diagonal. In this basis, we have further rotated the parameters $\rho_a^{}$ to be real, i.e. $\rho_a^{}=\rho_a^\ast$.

\section{Neutrino mass}

 \begin{figure*}
\centering
\includegraphics[scale=0.6]{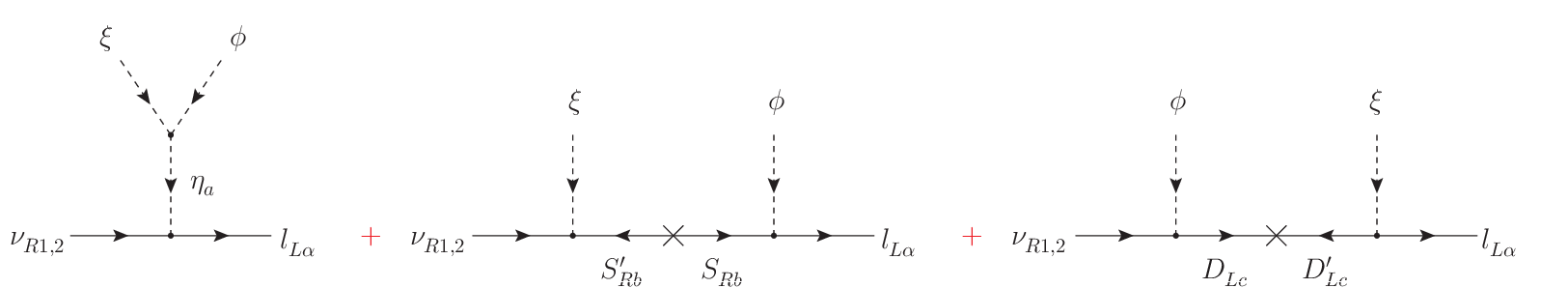} \caption{\label{numass} The Dirac neutrino mass generation.}
\end{figure*}

We can integrate out the heavy Higgs doublet(s) $\eta$, fermion singlet(s) $S$ and fermion doublet(s) $D$ to realize the effective operator (\ref{nmass}),
\begin{eqnarray}
\mathcal{L}&\supset& \sum_{i=1,2 \atop \alpha=e,\mu,\tau}^{a,b,c=1,...} \left(\frac{\rho_a^{} }{M_{\eta_a^{}}^2} f_{a\alpha i}^{} + y_{L\alpha b}^{}\frac{1}{M_{S_b^{}}^{}} y_{R b i }^{}\right.\nonumber\\
[1mm]
&&\left.+ g_{L\alpha c}^{}\frac{1}{M_{D_c^{}}^{}} g_{R c i }^{} \right)\bar{l}_{L\alpha}^{} \phi \nu_{Ri}^{}\xi  + \textrm{H.c.}\,.
\end{eqnarray}
When the Higgs singlet $\xi$ develops its VEV $\langle\xi\rangle$ for the $U(1)_{B-L}^{}$ symmetry breaking, the two right-handed neutrinos $\nu_{R1}^{}$ and $\nu_{R2}^{}$ can acquire the Yukawa couplings to the SM lepton and Higgs doublets $l_L^{}$ and $\phi$, i.e.
\begin{eqnarray}
\mathcal{L} &\supset& -\sum_{i=1,2\atop \alpha=e,\mu\,\tau}y_{\nu\alpha i}^{}\bar{l}_{L\alpha}^{} \phi \nu_{Ri}^{} ~~ \textrm{with}\nonumber\\
[2mm]
&& y_{\nu\alpha i}^{} = -\sum_{a,b,c=1,...}^{}\left(\frac{\rho_a^{} \langle \xi\rangle }{M_{\eta_a^{}}^2} f_{a \alpha i}^{} +y_{L\alpha b}^{}\frac{\langle\xi\rangle }{M_{S_b^{}}^{}} y_{R b i }^{} \right.\nonumber\\
[1mm]
&&\left.\quad\quad \quad
+g_{L\alpha b}^{}\frac{\langle\xi\rangle }{M_{D_c^{}}^{}} g_{R c i }^{}\right).
\end{eqnarray}
Therefore, we can obtain a Dirac neutrino mass matrix,
\begin{eqnarray}
\label{minimal}
\mathcal{L}\supset -\sum_{i=1,2 \atop \alpha=e,\mu,\tau}m_{\nu\alpha i}^{}\bar{\nu}_{L\alpha}^{} \nu_{Ri}^{} ~~\textrm{with}~~m_{\nu}^{} = y_{\nu}^{} \langle \phi\rangle \,.
\end{eqnarray}
The above Dirac neutrino mass generation can be also understood by Fig. \ref{numass}.

The experimental limit on the $U(1)_{B-L}^{}$ symmetry breaking scale is a few TeV, as shown in Eq. (\ref{low}). On the other hand, we will show later the third and forth right-handed neutrinos $\nu_{R3,4}^{}$ are expected to form a dark matter particle. To account for the observed dark matter relic density, the annihilations of the dark matter right-handed neutrinos into the light species should have a right cross section. The upper bound of the $U(1)_{B-L}^{}$ symmetry breaking scale thus should not be far above the TeV scale unless a fine-tuned resonant enhancement \cite{imy2009} is introduced in the s-channel dark matter annihilations. Therefore, the Dirac neutrino masses can be highly suppressed in a natural way as long as the masses of the heavy Higgs doublet(s) $\eta$, fermion singlet(s) $S$ and/or fermion doublet(s) $D$ are much larger than the TeV scale. For example, we can take
\begin{eqnarray}
 \langle \xi \rangle = \mathcal{O}(10\,\textrm{TeV})\,,
\end{eqnarray}
and then obtain  
\begin{eqnarray}
m_\nu^{}=\mathcal{O}\left(0.01-0.1\,\textrm{eV}\right) ~~\textrm{for}~~\langle \phi\rangle =174\,\textrm{GeV}\,,
\end{eqnarray}
by further inputting 
\begin{eqnarray}
&&M_{\eta_a^{}}^{}= \mathcal{O}\left(10^{15}_{}\,\textrm{GeV}\right)\,,~~\rho_a^{} = \mathcal{O}\left(10^{14}_{}\,\textrm{GeV}\right)\,,\nonumber\\
[1mm]
&&f_{a\alpha i}^{}=  \mathcal{O}\left(0.1\right)\,;\nonumber\\
[3mm]
&&M_{S_b^{}}^{} = \mathcal{O}\left(10^{15}_{}\,\textrm{GeV}\right)\,,~~y_{L\alpha b}^{}=  \mathcal{O}\left(0.1\right)\,,\nonumber\\
[1mm]
&&y_{R bi }^{}=  \mathcal{O}\left(0.1\right)\,;\nonumber\\
[3mm]
&&M_{D_c^{}}^{} = \mathcal{O}\left(10^{15}_{}\,\textrm{GeV}\right)\,,~~g_{L\alpha c}^{}=  \mathcal{O}\left(0.1\right)\,,\nonumber\\
[1mm]
&&g_{R ci }^{}=  \mathcal{O}\left(0.1\right)\,.
\end{eqnarray}

In Eq. (\ref{minimal}), the Dirac neutrino mass matrix only involves two right-handed neutrinos so that it can have at most two nonzero eigenvalues. This fact is independent on the number of the heavy Higgs doublet(s) $\eta$, fermion singlet(s) $S$ and/or fermion doublet(s) $D$. Since the current experimental data indicate the existence of at least two massive neutrinos \cite{navas2024}, we would like to name such $3\times 2$ neutrino mass matrix with rank 2 as a minimal Dirac neutrino mass matrix. Note if we do not introduce the heavy Higgs doublet(s) $\eta$, we should have at least two heavy fermion singlets $S$, or at least two heavy fermion doublets $D$, or at least one heavy fermion singlet $S$ and at least one heavy fermion doublet $D$ to guarantee the rank 2 and the two nonzero eigenvalues of the neutrino mass matrix.

\section{Baryon asymmetry}

We will show in the following that a successful leptogenesis needs (i) at least two heavy Higgs doublets $\eta$, or (ii) at least two heavy fermion singlets $S$, or (iii) at least two heavy fermion doublets $D$, or (iv) at least one heavy Higgs doublet $\eta$ and at least one heavy fermion singlet $S$, or (v) at least one heavy Higgs doublet $\eta$ and at least one heavy fermion doublet $D$, or (vi) at least one heavy fermion singlet $S$ and at least one heavy fermion doublet $D$.

\subsection{Heavy Higgs doublet decays}

 \begin{figure*}
\centering
\includegraphics[scale=0.6]{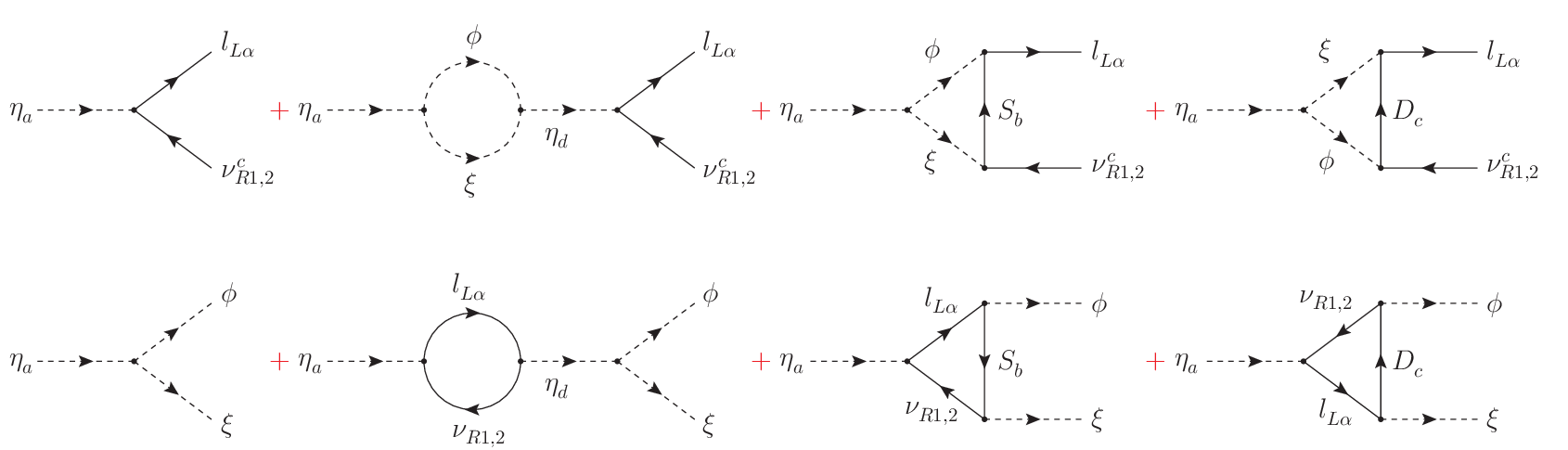} \caption{\label{hdecay} The lepton-number-conserving decays of the heavy Higgs doublets.}
\end{figure*}

As shown in Fig. \ref{hdecay}, there are two decay modes of the heavy Higgs doublet(s) $\eta$, i.e.
\begin{eqnarray}
\eta \rightarrow l_L^{}+ \nu_{R1,2}^{c}\,,~~ \eta \rightarrow \phi + \xi\,.
\end{eqnarray}
If the CP is not conserved, we can expect a CP asymmetry in the above decays,
\begin{eqnarray}
\varepsilon_{\eta_a^{}}^{}&=&\frac{\Gamma(\eta_a^{} \rightarrow l_L^{} +\nu_{R1,2}^{c})-\Gamma( \eta_a^{\ast} \rightarrow l_L^c +\nu_{R1,2}^{})}{\Gamma_{\eta_a^{}}^{}}\nonumber\\
[2mm]
&=&\frac{\Gamma( \eta^\ast_{a} \rightarrow \phi^\ast_{} +\xi^\ast_{} )-\Gamma( \eta_a^{} \rightarrow \phi + \xi)}{\Gamma_{\eta_a^{}}^{}}\neq 0\,,
\end{eqnarray}
where $\Gamma_{\eta_a^{}}^{}$ is the total decay width,
\begin{eqnarray}
\Gamma_{\eta_a^{}}^{}&=&\Gamma(\eta_a^{} \rightarrow l_L^{} +\nu_{R1,2}^{c})+\Gamma( \eta_a^{} \rightarrow \phi +\xi)\nonumber\\
[2mm]
&=&\Gamma(\eta^\ast_{a} \rightarrow l_L^{c} +\nu_{R1,2}^{})+\Gamma( \eta^\ast_{a} \rightarrow \phi^\ast_{}+ \xi^\ast_{} )\,.
\end{eqnarray}
We can calculate the decay width at tree level and the CP asymmetry at one-loop order,
\begin{eqnarray}
\Gamma_{\eta_a^{}}^{}&=&\frac{1}{16\pi}\left[\textrm{Tr}\left(f^\dagger_{a}f_a^{}\right)+ \frac{\rho_a^2}{M_{\eta_a^{}}^{2}}\right]M_{\eta_a^{}}^{}\,,
\end{eqnarray}
\begin{eqnarray}
\label{cpeta}
\varepsilon_{\eta_a^{}}^{}&=&-\frac{1}{4\pi}\left\{\sum_{d}^{}\frac{\textrm{Im}\left[\textrm{Tr}\left(f_a^\dagger f_d^{}\right)\right]}{\left(f^\dagger_{}f\right)_{aa}^{}+ \frac{\rho_a^2}{M_{\eta_a^{}}^{2}}}\right.\nonumber\\
[1mm]
&&\times \frac{\rho_a^{}\rho_d^{}\left(M_{\eta_d^{}}^2-M_{\eta_a^{}}^2\right)}{\left(M_{\eta_d^{}}^2-M_{\eta_a^{}}^2\right)^2_{}+M_{\eta_a^{}}^{2} \Gamma_{\eta_d^{}}^{2} }\nonumber\\
[1mm]
&&\left.+\sum_{b}^{}\frac{\textrm{Im}\left[\left(y_R^{}f_a^\dagger y_{L}^{}\right)_{bb}^{}\right]}{\left(f^\dagger_{}f\right)_{aa}^{}+ \frac{\rho_a^2}{M_{\eta_a^{}}^{2}}}\right.\nonumber\\
[1mm]
&&\times \frac{\rho_a^{}M_{S_b^{}}^{}}{M_{\eta_a^{}}^2}
\ln\left(1+\frac{M_{\eta_a^{}}^2}{M_{S_b^{}}^2}\right)\nonumber\\
[1mm]
&&\left.+\sum_{c}^{}\frac{\textrm{Im}\left[\left(g_R^{}f_a^\dagger g_{L}^{}\right)_{cc}^{}\right]}{\left(f^\dagger_{}f\right)_{aa}^{}+ \frac{\rho_a^2}{M_{\eta_a^{}}^{2}}}\right.\nonumber\\
[1mm]
&&\left.\times \frac{\rho_a^{}M_{D_c^{}}^{}}{M_{\eta_a^{}}^2}
\ln\left(1+\frac{M_{\eta_a^{}}^2}{M_{D_c^{}}^2}\right)\right\}\,.
\end{eqnarray}
Here the first term in the CP asymmetry is the self-energy correction mediated by the heavy Higgs doublet(s) while the second and third terms respectively are the vertex corrections mediated by the heavy fermion singlet(s) and fermion doublet(s). Clearly, a nonzero CP asymmetry $\varepsilon_{\eta_a^{}}^{}$ needs at least two heavy Higgs doublets $\eta$, or at least one heavy Higgs doublet $\eta$ and at least one heavy fermion singlet $S$, or at least one heavy Higgs doublet $\eta$ and at least one heavy fermion doublet $D$.

\subsection{Heavy fermion singlet decays}

 \begin{figure*}
\centering
\includegraphics[scale=0.6]{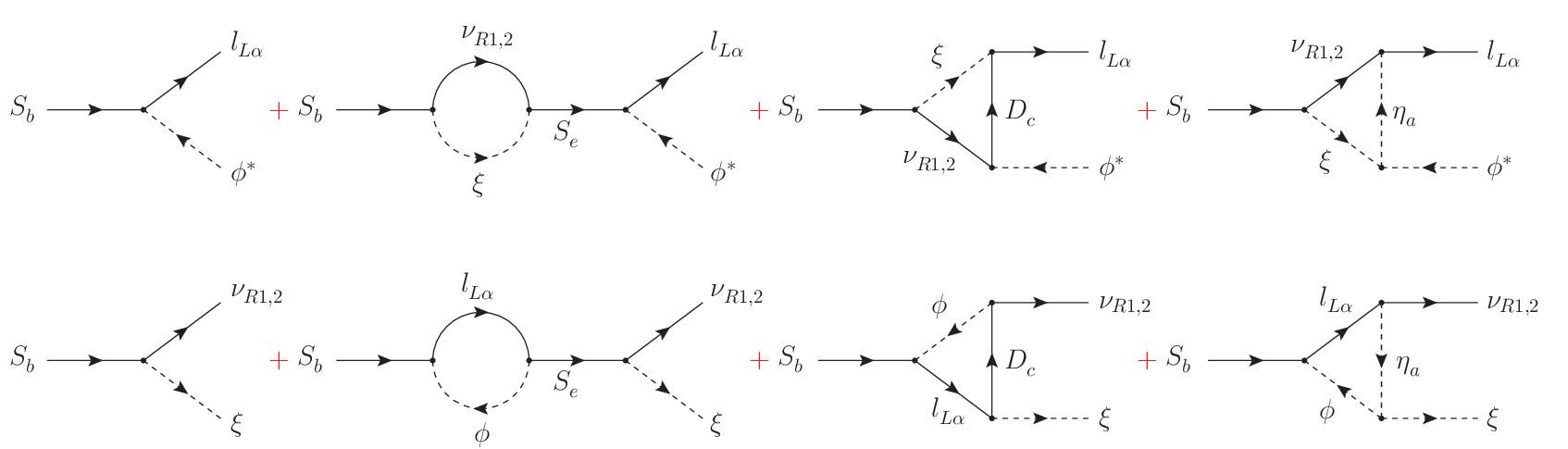} \caption{\label{sdecay} The lepton-number-conserving decays of the heavy fermion singlets.}
\end{figure*}

As for the heavy fermion singlet(s) $S$, their decay modes are 
\begin{eqnarray}
S_b^{} \rightarrow l_L^{} + \phi^\ast_{}\,,~~ S_b^{} \rightarrow \nu_{R1,2}^{}+ \xi \,.
\end{eqnarray}
The relevant diagrams are shown in Fig. \ref{sdecay}. The decay width and CP asymmetry can be calculated by 
\begin{eqnarray}
\Gamma_{S_b^{}}^{}&=& \Gamma(S_b^{} \rightarrow l_L^{}+ \phi^\ast_{})+\Gamma( S_b^{} \rightarrow \nu_{R1,2}^{} + \xi)\nonumber\\
[2mm]
&=& \Gamma(S_b^{c} \rightarrow l_L^{c} + \phi)+\Gamma( S_b^c \rightarrow \nu_{R1,2}^c + \xi^\ast_{})\nonumber\\
[2mm]
&=&\frac{1}{16\pi}\left[\left(y_L^\dagger y_L^{}\right)_{bb}^{}+\frac{1}{2}\left(y_R^{} y_R^\dagger\right)_{bb}^{}\right]M_{S_b^{}}^{}\,,
\end{eqnarray}
\begin{eqnarray}
\label{cps}
\varepsilon_{S_b^{}}^{}&=& \frac{\Gamma(S_b^{} \rightarrow l_L^{} + \phi^\ast_{})- \Gamma(S_b^{c} \rightarrow l_L^{c} + \phi)}{\Gamma_{S_b^{}}^{}}\nonumber\\
[2mm]
&=&\frac{\Gamma( S_b^c \rightarrow \nu_{R1,2}^c + \xi^\ast_{})- \Gamma( S_b^{} \rightarrow \nu_{R1,2}^{} + \xi)}{\Gamma_{S_b^{}}^{}}\nonumber\\
[2mm]
&=&\frac{1}{8\pi}\left\{\sum_{d}^{}\frac{\textrm{Im}\left[\left(y_{L}^\dagger y_{L}^{}\right)_{bd}\left(y_{R}^{} y_{R}^\dagger\right)_{db}^{}\right]}{\left(y_L^\dagger y_L^{}\right)_{bb}^{}+\frac{1}{2}\left(y_R^{} y_R^\dagger\right)_{bb}^{}}\right.\nonumber\\
[1mm]
&&\left.\times \frac{M_{S_b^{}}^{}M_{S_d^{}}^{}\left(M_{S_d^{}}^2- M_{S_b^{}}^2\right)}{\left(M_{S_d^{}}^2- M_{S_b^{}}^2\right)^2_{}+ M_{S_b^{}}^{2} \Gamma_{S_d^{}}^{2}}\right.\nonumber\\
[1mm]
&&\left.+\sum_{c}^{}\frac{\textrm{Im}\left[\left(y_{L}^\dagger g_{L}^{}\right)_{bc}^{}\left(g_R^{} y_R^\dagger\right)_{cb}^{}\right]}{\left(y_L^\dagger y_L^{}\right)_{bb}^{}+\frac{1}{2}\left(y_R^{} y_R^\dagger\right)_{bb}^{}}\right.\nonumber\\
[1mm]
&&\left.\times \frac{2M_{D_c}^{}}{M_{S_b}^{}}\left[1-\left(1+\frac{M_{D_c}^2}{M_{S_b}^2}\right)\ln \left(1+\frac{M_{S_b}^2}{M_{D_c}^2}\right)\right]\right.\nonumber\\
[1mm]
&&\left.+\sum_{a}^{}\frac{\textrm{Im}\left[\left(y_{L}^\dagger f_a^{}y_{R}^\dagger\right)_{bb}^{}\right]}{\left(y_L^\dagger y_L^{}\right)_{bb}^{}+\frac{1}{2}\left(y_R^{} y_R^\dagger\right)_{bb}^{}}\right.\nonumber\\
[1mm]
&&\left.\times \frac{2\rho_a^{}}{M_{S_b^{}}^{}}\left[-1+\frac{M_{\eta_a^{}}^2}{M_{S_b^{}}^2}\ln\left(1+\frac{M_{S_b^{}}^2}{M_{\eta_a^{}}^2}\right)\right]\right\}\,.
\end{eqnarray}
Here the first term in the CP asymmetry is the self-energy correction mediated by the heavy fermion singlet(s) while the second and third terms are the vertex correction mediated by the heavy fermion doublet(s) and Higgs doublet(s). It is easy to check that the CP asymmetry $\varepsilon_{S_b^{}}^{}$ can arrive at a nonzero value if and only if the models contain at least two heavy fermion singlets $S$, or at least one heavy fermion singlet $S$ and at least one heavy fermion doublet $D$, or at least one heavy fermion singlet $S$ and at least one heavy Higgs doublet $\eta$.

\subsection{Heavy fermion doublet decays}

 \begin{figure*}
\centering
\includegraphics[scale=0.6]{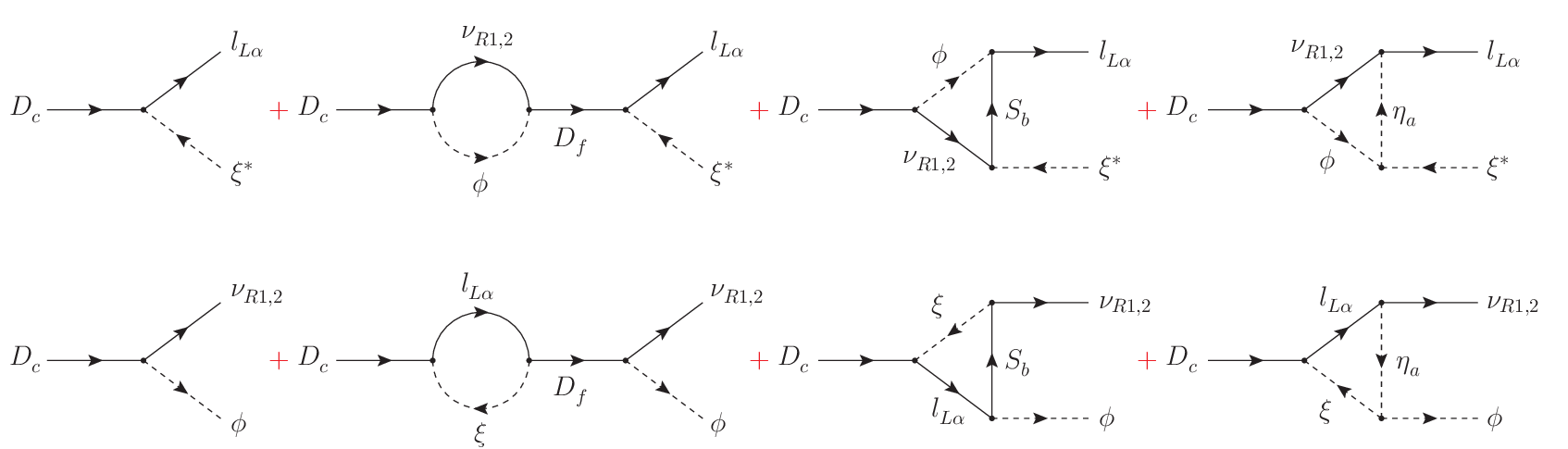} \caption{\label{ddecay}  The lepton-number-conserving decays of the heavy fermion doublets.}
\end{figure*}

We also consider the decays of the heavy fermion doublet(s) $D$, i.e. 
\begin{eqnarray}
D_c^{} \rightarrow l_L^{} + \xi^\ast_{}\,,~~ D_c^{} \rightarrow \nu_{R1,2}^{} + \phi \,.
\end{eqnarray}
See the relevant diagrams in Fig. \ref{ddecay}. The decay width and CP asymmetry can be calculated by 
\begin{eqnarray}
\Gamma_{D_c^{}}^{}&=& \Gamma(D_c^{} \rightarrow l_L^{} \phi^\ast_{})+\Gamma( D_c^{} \rightarrow \nu_{R1,2}^{} \xi)\nonumber\\
[2mm]
&=& \Gamma(D_c^{c} \rightarrow l_L^{c} \phi)+\Gamma( D_c^c \rightarrow \nu_{R1,2}^c \xi^\ast_{})\nonumber\\
[2mm]
&=&\frac{1}{32\pi}\left[\left(g_L^\dagger g_L^{}\right)_{cc}^{}+\left(g_R^{} g_R^\dagger\right)_{cc}^{}\right]M_{N_b^{}}^{}\,,
\end{eqnarray}
\begin{eqnarray}
\label{cpd}
\varepsilon_{D_c^{}}^{}&=& \frac{\Gamma(D_c^{} \rightarrow l_L^{} \phi^\ast_{})- \Gamma(D_c^{c} \rightarrow l_L^{c} \phi)}{\Gamma_{N_b^{}}^{}}\nonumber\\
[2mm]
&=&\frac{\Gamma( D_c^c \rightarrow \nu_{R1,2}^c \xi^\ast_{})- \Gamma( D_c^{} \rightarrow \nu_{R1,2}^{} \xi)}{\Gamma_{N_b^{}}^{}}\nonumber\\
[2mm]
&=&\frac{1}{8\pi}\left\{\sum_{d}^{}\frac{\textrm{Im}\left[\left(g_{L}^\dagger g_{L}^{}\right)_{cd}\left(g_{R}^{} g_{R}^\dagger\right)_{dc}^{}\right]}{\left(g_L^\dagger g_L^{}\right)_{cc}^{}+\left(g_R^{} g_R^\dagger\right)_{cc}^{}}\right.\nonumber\\
[1mm]
&&\left.\times \frac{M_{D_c^{}}^{}M_{D_d^{}}^{}\left(M_{D_d^{}}^2- M_{D_c^{}}^2\right)}{\left(M_{D_d^{}}^2- M_{D_c^{}}^2\right)^2_{} + M_{D_c^{}}^{2}\Gamma_{D_d^{}}^{2}}\right.\nonumber\\
[1mm]
&&\left.+\sum_{b}^{}\frac{\textrm{Im}\left[\left(g_{L}^\dagger y_{L}^{}\right)_{cb}^{}\left(y_R^{} g_R^\dagger\right)_{bc}^{}\right]}{\left(g_L^\dagger g_L^{}\right)_{cc}^{}+\left(g_R^{} g_R^\dagger\right)_{cc}^{}}\right.\nonumber\\
[1mm]
&&\left.\times \frac{2M_{S_b}^{}}{M_{D_c}^{}}\left[1-\left(1+\frac{M_{S_b}^2}{M_{D_c}^2}\right)\ln \left(1+\frac{M_{D_c}^2}{M_{S_b}^2}\right)\right]\right.\nonumber\\\nonumber\\
[1mm]
&&\left.+\sum_{a}^{}\frac{\textrm{Im}\left[\left(g_{L}^\dagger f_a^{}g_{R}^\dagger\right)_{cc}^{}\right]}{\left(g_L^\dagger g_L^{}\right)_{cc}^{}+\left(g_R^{} g_R^\dagger\right)_{cc}^{}}\right.\nonumber\\
[1mm]
&&\left.\times \frac{2\rho_a^{}}{M_{N_b^{}}^{}}\left[-1+\frac{M_{\eta_a^{}}^2}{M_{N_b^{}}^2}\ln\left(1+\frac{M_{N_b^{}}^2}{M_{\eta_a^{}}^2}\right)\right]\right\}\,.
\end{eqnarray}
Here the first term in the CP asymmetry is the self-energy correction mediated by the heavy fermion doublet(s) while the second and third terms are the vertex correction mediated by the heavy fermion singlet(s) and Higgs doublet(s). Obviously, a nonzero CP asymmetry $\varepsilon_{D_c^{}}^{}$ requires the existence of at least two heavy fermion doublets $D$, or at least one heavy fermion doublet $D$ and at least one heavy fermion singlet $S$, or at least one heavy fermion doublet $D$ and at least one heavy Higgs doublet $\eta$.

\subsection{Final baryon asymmetry}

When the heavy Higgs doublets $\eta_a^{}$, the heavy fermion singlets $S_b^{}$ and/or the fermion doublets $D_c^{}$ go out of equilibrium, their decays can generate a lepton number $L_{l_L^{}}^{}$ stored in the SM lepton doublets $l_L^{}$ and an opposite lepton number $L_{\nu_{R1,2}^{}+\xi}^{}$ stored in the right-handed neutrinos $\nu_{R1,2}^{}$ and the Higgs singlet $\xi$.

For example, if the Higgs doublet $\eta_1^{}$ is much lighter than the other heavy Higgs doublet(s) $\eta_{a\neq 1}^{}$, the heavy fermion singlet(s) $S_b^{}$ and the heavy fermion doublet(s) $D_c^{}$, its decays will dominate the final lepton numbers \cite{kt1990},
\begin{eqnarray}
\label{hhl}
L_{l_L^{}}^{}=-L_{\nu_{R1,2}^{}+\xi}^{} = \varepsilon_{\eta_1^{}}^{}\left(\frac{n^{eq}_{\eta_1^{}} }{s}\right)\left|_{T=T_D^{}}^{}\right.\,.
\end{eqnarray}
Here the CP asymmetry $ \varepsilon_{\eta_1^{}}^{}$ has an upper bound \cite{di2002}
\begin{eqnarray}
\label{cphh1}
\varepsilon_{\eta_1^{}}^{}&\simeq&\frac{1}{4\pi}\frac{\textrm{Im}\left[\textrm{Tr}\left(f_1^\dagger m_\nu^{}\right)\right]\rho_1^{}}{\left[\left(f^\dagger_{}f\right)_{11}^{}+ \frac{\rho_1^2}{M_{\eta_1^{}}^{2}}\right]\langle\phi\rangle \langle\xi\rangle}\nonumber\\
[1mm]
&\leq&\frac{1}{8\pi}\frac{\textrm{Im}\left[\textrm{Tr}\left(f_1^\dagger m_\nu^{}\right)\right]\rho_1^{}}{\sqrt{\left(f^\dagger_{}f\right)_{11}^{} \frac{\rho_1^2}{M_{\eta_1^{}}^{2}}}\langle\phi\rangle \langle\xi\rangle}\nonumber\\
[1mm]
&<& \frac{1}{8\pi}\frac{\textrm{Tr}\left(f_1^\dagger\right) \rho_1^{}  m_{\textrm{max}}^{}}{\sqrt{\left(f^\dagger_{}f\right)_{11}^{} \frac{\rho_1^2}{M_{\eta_1^{}}^{}}  }\langle\phi\rangle \langle\xi\rangle}\nonumber\\
[1mm]
&=& \frac{1}{8\pi} \frac{m_{\textrm{max}}^{} M_{\eta_1^{}}^{}}{\langle\phi\rangle \langle\xi\rangle}\sin\delta_\eta^{}~~\textrm{with}\nonumber\\
[1mm]
&& \sin\delta_\eta^{} =\frac{\textrm{Tr}\left(f_1^\dagger\right) }{\sqrt{\left(f^\dagger_{}f\right)_{11}^{}}} \,.
\end{eqnarray}
Here and thereafter $m_{\textrm{max}}^{}$ denotes the largest eigenvalue of the neutrino mass matrix $m_\nu^{}$.

We can also consider the limiting case where the fermion singlet $S_1^{}$ is much lighter than the other heavy fermion singlet(s) $S_{b\neq 1}^{}$, the heavy Higgs doublet(s) $\eta_a^{}$ and the heavy fermion doublet(s) $D_c^{}$. The final lepton numbers then should be \cite{kt1990},
\begin{eqnarray}
\label{hfsl}
L_{l_L^{}}^{}=-L_{\nu_{R1,2}^{}+\xi}^{} = \varepsilon_{S_1^{}}^{}\left(\frac{n^{eq}_{S_1^{}} }{s}\right)\left|_{T=T_D^{}}^{}\right.\,,
\end{eqnarray}
where the CP asymmetry $ \varepsilon_{S_1^{}}^{}$ has an upper bound \cite{di2002}
\begin{eqnarray}
\label{cphfs1}
\varepsilon_{S_1^{}}^{}&\simeq&\frac{1}{8\pi}\frac{\textrm{Im}\left[\left(y_{L}^\dagger m_\nu^{} y_{R}^\dagger\right)_{11}^{}\right]M_{S_1^{}}^{}}{\left[\left(y_L^\dagger y_L^{}\right)_{11}^{}+\frac{1}{2}\left(y_R^{} y_R^\dagger\right)_{11}^{}\right]\langle\phi\rangle \langle\xi\rangle}\nonumber\\
[1mm]
&\leq& \frac{1}{8\pi}\frac{\textrm{Im}\left[\left(y_{L}^\dagger m_\nu^{} y_{R}^\dagger\right)_{11}^{}\right]M_{S_1^{}}^{}}{\sqrt{2\left(y_L^\dagger y_L^{}\right)_{11}^{}\left(y_R^{} y_R^\dagger\right)_{11}^{}}\langle\phi\rangle \langle\xi\rangle}\nonumber\\
[1mm]
&<& \frac{1}{8\pi}\frac{\left(y_{L}^\dagger y_{R}^\dagger\right)_{11}^{} m_{\textrm{max}}^{} M_{S_1^{}}^{}}{\sqrt{2\left(y_L^\dagger y_L^{}\right)_{11}^{}\left(y_R^{} y_R^\dagger\right)_{11}^{}}\langle\phi\rangle \langle\xi\rangle}\nonumber\\
[1mm]
&= &  \frac{1}{8\pi}\frac{m_{\textrm{max}}^{} M_{N_1^{}}^{}}{\langle\phi\rangle \langle\xi\rangle}\sin\delta_S^{}~~\textrm{with}\nonumber\\
[1mm]
&&\sin\delta_S^{}= \frac{\left(y_{L}^\dagger y_{R}^\dagger\right)_{11}^{}}{\sqrt{2\left(y_L^\dagger y_L^{}\right)_{11}^{}\left(y_R^{} y_R^\dagger\right)_{11}^{}}}\,.
\end{eqnarray}

Alternatively, we can consider another simple case where the fermion doublet $D_1^{}$ is much lighter than the other heavy fermion singlet(s) $D_{c\neq 1}^{}$, the heavy Higgs doublet(s) $\eta_a^{}$ and the heavy fermion singlet(s) $N_{b}^{}$. The final lepton numbers then should be \cite{kt1990},
\begin{eqnarray}
\label{hfdl}
L_{l_L^{}}^{}=-L_{\nu_{R1,2}^{}+\xi}^{} = \varepsilon_{D_1^{}}^{}\left(\frac{n^{eq}_{D_1^{}} }{s}\right)\left|_{T=T_D^{}}^{}\right.\,,
\end{eqnarray}
where the CP asymmetry $ \varepsilon_{D_1^{}}^{}$ has an upper bound \cite{di2002} 
\begin{eqnarray}
\label{cphfd1}
\varepsilon_{D_1^{}}^{}&\simeq&\frac{1}{4\pi}\frac{\textrm{Im}\left[\left(g_{L}^\dagger m_\nu^{} g_{R}^\dagger\right)_{11}^{}\right]M_{D_1^{}}^{}}{\left[\left(g_L^\dagger g_L^{}\right)_{11}^{}+\left(g_R^{} g_R^\dagger\right)_{11}^{}\right]\langle\phi\rangle \langle\xi\rangle}\nonumber\\
[1mm]
&\leq& \frac{1}{8\pi}\frac{\textrm{Im}\left[\left(g_{L}^\dagger m_\nu^{} g_{R}^\dagger\right)_{11}^{}\right]M_{D_1^{}}^{}}{\sqrt{\left(g_L^\dagger g_L^{}\right)_{11}^{}\left(g_R^{} g_R^\dagger\right)_{11}^{}}\langle\phi\rangle \langle\xi\rangle}\nonumber\\
[1mm]
&<& \frac{1}{8\pi}\frac{\left(g_{L}^\dagger g_{R}^\dagger\right)_{11}^{} m_{\textrm{max}}^{} M_{D_1^{}}^{}}{\sqrt{\left(g_L^\dagger g_L^{}\right)_{11}^{}\left(g_R^{} g_R^\dagger\right)_{11}^{}}\langle\phi\rangle \langle\xi\rangle}\nonumber\\
[1mm]
&=&  \frac{1}{8\pi}\frac{m_{\textrm{max}}^{} M_{D_1^{}}^{}}{\langle\phi\rangle \langle\xi\rangle} \sin\delta_D^{} ~~\textrm{with}\nonumber\\
[1mm]
&&\sin\delta_D^{} = \frac{\left(g_{L}^\dagger g_{R}^\dagger\right)_{11}^{} }{\sqrt{\left(g_L^\dagger g_L^{}\right)_{11}^{}\left(g_R^{} g_R^\dagger\right)_{11}^{}}}\,.
\end{eqnarray}

In Eqs. (\ref{hhl}), (\ref{hfsl}) and (\ref{hfdl}), $n_{\eta_1^{},S_1^{},D_1^{}}^{eq}$ and $T_D^{}$ respectively are the equilibrium number density and the decoupled temperature of the decaying heavy particles, while $s$ is the entropy density of the universe. The decay-produced lepton number in the SM lepton doublets can be partially converted to a baryon asymmetry by the sphaleron processes \cite{ht1990}, 
\begin{eqnarray}
\label{sph}
B= -\frac{28}{79} L_{l_L^{}}^{}\,.
\end{eqnarray}

In the weak washout region where \cite{kt1990}, 
\begin{eqnarray}
\label{weak}
\left.K =\frac{\Gamma_{\eta_1^{}/S_1^{}/D_1^{}}^{}}{2H(T)}\right|_{T=M_{\eta_1^{}/S_1^{}/D_1^{}}}^{}< 1\,,
\end{eqnarray}
we can approximately obtain the lepton numbers (\ref{hhl}), (\ref{hfsl}) and (\ref{hfdl}) by  
\begin{eqnarray}
L_{l_L^{}}^{}=-L_{\nu_{R1,2}^{}+\xi}^{} \sim  \frac{\varepsilon_{\eta_1^{}/S_1^{}/D_1^{}}^{}}{g_\ast^{}}\,.
\end{eqnarray} 
Here
\begin{eqnarray}
 H(T)&=&\left(\frac{8\pi^{3}_{}g_{\ast}^{}}{90}\right)^{\frac{1}{2}}_{}\frac{T^2_{}}{M_{\textrm{Pl}}^{}}\,,
 \end{eqnarray}
is the Hubble constant with $M_{\textrm{Pl}}^{}\simeq 1.22\times 10^{19}_{}\,\textrm{GeV}$ being the Planck mass and $g_{\ast}^{}=117.75$ being the relativistic degrees of freedom (the SM fields plus the right-handed neutrinos $\nu_{R1,2,3,4}^{}$, the Higgs singlet $\xi$ and the $U(1)_{B-L}^{}$ gauge field $Z_{B-L}^{}$.). The baryon number (\ref{sph}) then can be given by 
\begin{eqnarray}
\label{bau}
B \sim  -\frac{28}{79} \frac{\varepsilon_{\eta_1^{}/S_1^{}/D_1^{}}^{}}{g_\ast^{}}\,.
\end{eqnarray}

For a numerical estimation, we can take  
\begin{eqnarray}
&&M_{\eta_1^{}}^{}=10^{15}_{}\,\textrm{GeV}\,,~~\rho_1^{} =10^{14}_{}\,\textrm{GeV}\,,
f_{1\alpha i}^{}=  \mathcal{O}\left(0.1\right)\,;\nonumber\\
&&\textrm{or}\nonumber\\
&&M_{S_1^{}}^{} = 10^{15}_{}\,\textrm{GeV} \,,~~y_{L\alpha 1}^{}=  \mathcal{O}\left(0.1\right)\,,~~ y_{R 1 i }^{}=  \mathcal{O}\left(0.1\right)\,;\nonumber\\
&&\textrm{or}\nonumber\\
&&M_{D_1^{}}^{} = 10^{15}_{}\,\textrm{GeV} \,,~~g_{L\alpha 1}^{}=  \mathcal{O}\left(0.1\right)\,,~~ g_{R 1 i}^{}=  \mathcal{O}\left(0.1\right)\,,\nonumber\\
&&
\end{eqnarray}
to fulfil the weak washout condition (\ref{weak}), i.e.
\begin{eqnarray}
 K = \mathcal{O}\left(0.1-1\right)
\end{eqnarray}
By further fixing 
\begin{eqnarray}
 \langle \xi \rangle = 20\,\textrm{TeV}\,,
\end{eqnarray}
we can read the CP asymmetry (\ref{cphh1}), (\ref{cphfs1}) or (\ref{cphfd1}) to be 
\begin{eqnarray}
\varepsilon_{\eta_1^{}/S_1^{}/D_1^{}}^{}< 5.7\times 10^{-5}_{} \left(\frac{m_{\textrm{max}}^{}}{0.05\,\textrm{eV}} \right) \left(\frac{\sin\delta_{\eta/S/D}^{}}{0.1}\right)\,.~~
\end{eqnarray}
This means the final baryon asymmetry (\ref{bau}) can explain the observation \cite{navas2024} as the CP asymmetry (\ref{cphh1}), (\ref{cphfs1}) or (\ref{cphfd1}) acquires a desired value, i.e. 
\begin{eqnarray}
B \sim  8.7 \times 10^{-11}_{}\left( \frac{\varepsilon_{\eta_1^{}/S_1^{}/D_1^{}}^{}}{-2.9\times 10^{-8}_{}}\right)\,.
\end{eqnarray}

\section{Dark matter and right-handed neutrinos}

Due to the Yukawa interaction (\ref{dm}), the third and forth right-handed neutrinos $\nu_{R3,4}^{}$ can form a Dirac particle after the $U(1)_{B-L}^{}$ symmetry breaking, i.e.
\begin{eqnarray}
\mathcal{L} &\supset&  i \bar{\chi}\gamma^\mu_{}\partial_\mu^{} \chi - m_\chi^{} \bar{\chi}\chi \nonumber\\
&& \textrm{with}~~ \chi = \nu_{R3}^{}+\nu_{R4}^{c}\,,~~ m_\chi^{}= y_{34}^{} \langle\xi\rangle\,.
 \end{eqnarray}
Clearly, the Dirac fermion $\chi$ will keep stable to leave a dark matter relic density. The dark matter annihilation and scattering can be determined by the related gauge and Yukawa interactions,
\begin{eqnarray}
\mathcal{L} &\supset&  \frac{1}{2} g_{B-L}^{}Z^\mu_{B-L} \bar{\chi}\gamma_\mu^{}\left(\sqrt{17}+\gamma_5^{}\right)\chi  - \frac{1}{\sqrt{2}}y_{34}^{} h_\xi^{} \bar{\chi}\chi\,,\nonumber\\
&&
 \end{eqnarray}
where $h_\xi^{}$ is the Higgs boson from the Higgs scalar $\xi$. The gauge boson $Z_{B-L}^{}$ also couples to the SM fermions as well as the first and second right-handed neutrinos $\nu_{R1,2}^{}$, 
\begin{eqnarray}
\mathcal{L} &\supset& g_{B-L}^{} Z_{B-L}^\mu \left[\sum_{i=1}^{3}\left(\frac{1}{3}\bar{d}_i^{}\gamma_\mu^{} d_i^{} + \frac{1}{3}\bar{u}_{i}^{}\gamma_\mu^{}u_{i}^{}-\bar{e}_{i}^{}\gamma_\mu^{} e_{i}^{} \right. \right.\nonumber\\
[2mm]
&&\left.\left.-\bar{\nu}_{Li}\gamma_\mu^{} \nu_{Li}^{}\right)-2 \bar{\nu}_{R1}^{}\gamma_\mu^{}\nu_{R1}^{} - 2 \bar{\nu}_{R2}^{}\gamma_\mu^{}\nu_{R2}^{}\right]\,.
\end{eqnarray}
The perturbation requirement then should put an upper bound on the gauge coupling $g_{B-L}^{}$, i.e. 
\begin{eqnarray}
\label{gbl}
\frac{\sqrt{17}}{2}g_{B-L}^{} < \sqrt{4\pi} \Rightarrow g_{B-L}^{}<  \sqrt{\frac{16\,\pi}{17}}\,.
\end{eqnarray} 
As for the Higgs boson $h_\xi^{}$, it can interact with the SM through a Higgs portal as below,
\begin{eqnarray}
\mathcal{L} &\supset& - \lambda_{\phi\xi}^{} \phi^\dagger_{}\phi \xi^\dagger_{}\xi \,.
 \end{eqnarray}

For demonstration, we shall focus on the case that the $U(1)_{B-L}^{}$ gauge interactions dominate the dark matter annihilations and scatterings \cite{amr2014}. In this case, the thermally averaging dark matter annihilating cross section is given by \cite{bhkk2009} 
\begin{eqnarray}
\label{ann}
\langle\sigma_{\textrm{A}}^{} v_{\textrm{rel}}^{} \rangle&=& \sum_{f=d_i^{},u_i^{},e_i^{},\nu_{Li}^{},\nu_{R1,2}^{} \atop i=1,2,3}^{}\langle\sigma(\chi+\chi^c_{}\rightarrow f+f^c_{}) v_{\textrm{rel}}^{}\rangle \nonumber\\
[2mm]
&\simeq &\frac{357g_{B-L}^4}{8\pi} \frac{m_\chi^2}{M_{Z_{B-L}^{}}^4} \nonumber\\
[2mm]
&=& \frac{357}{32\pi} \frac{m_\chi^2}{\langle\xi\rangle^4_{}} =  \frac{327}{32\pi} \frac{y_{34}^2}{\langle\xi\rangle^2_{}}  \,,
\end{eqnarray}
where not only three generations of the SM fermions \cite{amr2014} but also two right-handed neutrinos $\nu_{R1,2}^{}$ are involved in the final states. The dark matter relic density then can well approximate to \cite{navas2024}
\begin{eqnarray}
\label{relic}
\Omega_\chi^{} h^2 \simeq \frac{0.1\,\textrm{pb}}{\langle \sigma_{\textrm{A}}^{} v_{\textrm{rel}}^{} \rangle} &=&0.1\,\textrm{pb}\times \frac{32\pi\langle\xi\rangle^4_{}}{357 m_{\chi}^2}\nonumber\\
[2mm]
&=&0.1\,\textrm{pb}\times \frac{32\pi \langle\xi\rangle^2_{}}{357 y_{34}^2} \,.
\end{eqnarray}
It should be noted that Eqs. (\ref{ann}) and (\ref{relic}) are based on the assumption, 
\begin{eqnarray}
4m_\chi^2 \ll M_{Z_{B-L}^{}}^2 \Rightarrow y_{34}^{2}\ll \frac{1}{2} g_{B-L}^{2} \,.
\end{eqnarray}
This indeed means 
\begin{eqnarray}
\label{y34}
y_{34}^{2}\ll \frac{8\pi}{17}   \Rightarrow  y_{34}^{}< \sqrt{\frac{8\pi}{17}} \,,
\end{eqnarray}
in the presence of the perturbation condition (\ref{gbl}).

By inserting the upper bound (\ref{y34}) into Eq. (\ref{relic}), we can put a constraint on the the VEV $\langle\xi\rangle$, i.e.
\begin{eqnarray}
\langle\xi\rangle &\simeq& \left(\frac{357 y_{34}^2 \Omega_\chi^{} h^2}{32\pi \times  0.1\,\textrm{pb}} \right)^{\frac{1}{2}}_{}\nonumber\\
[2mm]
&=&  47\,\textrm{TeV}\left(\frac{y_{34}^{}}{\sqrt{8\pi/17}}\right) \left(\frac{\Omega_\chi^{} h^2}{0.11}\right)^{\frac{1}{2}}_{}\nonumber\\
[2mm]
&<&47\,\textrm{TeV}\left(\frac{\Omega_\chi^{} h^2}{0.11}\right)^{\frac{1}{2}}_{}\,.
\end{eqnarray}
besides the experimental limit (\ref{low}). The dark matter mass, 
\begin{eqnarray}
\label{dm1}
m_\chi^{} &\simeq &\left(0.1\,\textrm{pb}\times \frac{32\pi\langle\xi\rangle^4_{}}{357\Omega_\chi^{} h^2} \right)^{\frac{1}{2}}_{} \nonumber\\
[2mm]
&=& 7\,\textrm{TeV}\left(\frac{\langle \xi\rangle }{16.5\,\textrm{TeV}}\right)^2_{}\left(\frac{0.11}{\Omega_\chi^{} h^2}\right)^{\frac{1}{2}}_{} \,,
\end{eqnarray}
thus should be in the range, 
\begin{eqnarray}
\label{dm2}
&&640\,\textrm{GeV}\left(\frac{0.11}{\Omega_\chi^{} h^2}\right)^{\frac{1}{2}}_{}\lesssim
m_\chi^{} < 57\,\textrm{TeV}\left(\frac{0.11}{\Omega_\chi^{} h^2}\right)^{\frac{1}{2}}_{}\nonumber\\
[2mm]
&&\textrm{for}~~5\,\textrm{TeV} \lesssim \langle\xi\rangle < 47\,\textrm{TeV}\,.
\end{eqnarray}

The gauge interactions can also mediate the dark matter scattering off nucleons. The dominant scattering cross section is spin independent \cite{jkg1996},
\begin{eqnarray}
\sigma_{\chi N}^{}&=& \frac{17 \,g_{B-L}^4}{4 \pi} \frac{\mu_r^2 }{M_{Z_{B-L}^{}}^4} \nonumber\\
[2mm]
&=& \frac{17 }{16\pi} \frac{\mu_r^2 }{\langle\xi\rangle^4_{}} =   \frac{2 }{21} \frac{\mu_r^2 }{m_\chi^2} \frac{0.1\,\textrm{pb}}{\Omega_\chi^{} h^2_{} }\,.
\end{eqnarray}
Here $\mu_r^{}=m_N^{}m_\chi^{}/(m_N^{}+m_\chi^{})$ is a reduced mass with $m_N^{}$ being the nucleon mass. As the dark matter is much heavier than the nucleon, the above dark matter scattering cross section indeed should be inversely proportional to the squared dark matter mass, 
\begin{eqnarray}
\sigma_{\chi N}^{}&=& 1.6\times 10^{-45}_{}\,\textrm{cm}^2_{} \left( \frac{7\,\textrm{TeV}}{m_\chi^{}}\right)^2_{} \nonumber\\
[1mm]
&&\times\left(\frac{\mu_r^{}}{940\,\textrm{MeV}}\right)^2_{} \left(\frac{0.11}{\Omega_\chi^{} h^2_{}}\right)\,.
\end{eqnarray}
By taking the dark matter direct detection results \cite{cui2017,aprile2018,meng2021,bo2024} into account, we can put a more stringent low limit on the dark matter mass \footnote{We may replace the $U(1)_{B-L}^{}$ gauge symmetry by a $U(1)_X^{}$ gauge symmetry where the $X$ number contains an arbitrary fraction of hypercharge besides the baryon-minus-lepton number \cite{langacker2009}. Consequently we may realize an isospin-violating dark matter \cite{fk2011} to relax the constraints from the dark matter direct detection experiments.},
\begin{eqnarray}
\label{dm3}
m_\chi^{} \gtrsim 7\,\textrm{TeV}\,.
\end{eqnarray}
So, the range (\ref{dm2}) should be modified by
\begin{eqnarray}
&&7\,\textrm{TeV}\left(\frac{0.11}{\Omega_\chi^{} h^2}\right)^{\frac{1}{2}}_{}\lesssim
m_\chi^{} < 57\,\textrm{TeV}\left(\frac{0.11}{\Omega_\chi^{} h^2}\right)^{\frac{1}{2}}_{}\nonumber\\
[2mm]
&&\textrm{for}~~16.5\,\textrm{TeV} \lesssim \langle\xi\rangle < 47\,\textrm{TeV}\,.
\end{eqnarray}

We also check if the right-handed neutrinos $\nu_{R1}^{}$ and $\nu_{R2}^{}$ can decouple above the QCD scale to satisfy the BBN constraint on the effective neutrino number. For this purpose, we need consider the annihilations of the right-handed neutrinos into the relativistic species at the QCD scale,   
\begin{eqnarray}
\sigma_{\nu_R^{}}^{} &=&\sum_{f=d,u,s,e,\mu,\nu_L^{}}^{}\sigma(\nu_R^{}+\nu_R^c\rightarrow f+f^c) \nonumber\\
[2mm]
&=& \frac{3g_{B-L}^4}{2\pi}\frac{s}{M_{Z_{B-L}}^4} =  \frac{3}{8\pi}\frac{s}{\langle\xi\rangle^4_{}}\,,
\end{eqnarray}
with $s$ being the Mandelstam variable. The interaction rate then should be \cite{gnrrs2003}
\begin{eqnarray}
\Gamma_{\nu_R^{}}^{} &=&\frac{\frac{T}{32\pi^4_{}}\int^{\infty}_{0} s^{3/2}_{} K_1^{}\left(\frac{\sqrt{s}}{T}\right) \sigma_{{\nu_R^{}}}^{}ds }{\frac{2}{\pi^2_{}}T^3_{}}= \frac{9}{2\pi^3_{}} \frac{T^5_{}}{\langle\xi\rangle^4_{}}\,,~~
\end{eqnarray}
with $K_1^{}$ being a Bessel function. We take $g_\ast^{}(300\,\textrm{MeV})\simeq 61.75$ and then find 
\begin{eqnarray}
\left[\Gamma_{\nu_R^{}}^{} < H(T)\right]_{T\gtrsim 300\,\textrm{MeV}}^{}~~\textrm{for}~~\langle\xi\rangle \gtrsim 8\,\textrm{TeV}\,.
\end{eqnarray}

\section{Conclusion}

In this paper we have shown a $U(1)_{B-L}^{}$ gauge symmetry can predict the existence of the Dirac neutrinos and the stable dark matter. Specifically, we have extended the SM $SU(3)_c^{}\times SU(2)_L^{} \times U(1)_Y^{}$ gauge symmetries by a $U(1)_{B-L}^{}$ gauge symmetry. We then introduced four right-handed neutrinos in order to cancel the gauge anomalies. Because of their Yukawa couplings to the Higgs singlet for spontaneously breaking the $U(1)_{B-L}^{}$ symmetry, two right-handed neutrinos can form a stable Dirac fermion and hence can account for the dark matter relic density.  Furthermore, mediated by additionally heavy Higgs doublet(s), fermion singlet(s) and/or fermion doublet(s), the other two right-handed neutrinos can have a dimension-5 operator with the SM lepton and Higgs doublets as well as the $U(1)_{B-L}^{}$ Higgs singlet. We hence can obtain a minimal Dirac neutrino mass matrix only with two nonzero eigenvalues. Finally, the interactions for the neutrino mass generation can also allow the decays of the heavy Higgs doublet(s), fermion singlet(s) and/or fermion doublet(s) to produce a lepton asymmetry motivated by the leptogenesis mechanism.

\textbf{Acknowledgement}: This work was supported in part by the National Natural Science Foundation of China under Grant No. 12175038 and 11675100, and in part by the Fundamental Research Funds for the Central Universities.

\appendix

\section{The $U(1)_{B-L}^{}$ gauge anomalies}

The $SU(3)_c^{}-SU(3)_c^{}-U(1)_{B-L}^{}$ anomaly is
\begin{eqnarray}
3\times 3\times \left[2\times \left(+\frac{1}{3}\right) -\left(+ \frac{1}{3}\right) - \left(+\frac{1}{3}\right) \right]=0\,.
\end{eqnarray}

The $SU(2)_L^{}-SU(2)_L^{}-U(1)_{B-L}^{}$ anomaly is 
\begin{eqnarray}
3\times 2 \times  \left[3\times\left(+ \frac{1}{3}\right) +\left(-1\right)\right]=0\,.
\end{eqnarray}

The $U(1)_Y^{}-U(1)_Y^{}-U(1)_{B-L}^{}$ anomaly is 
\begin{eqnarray}
\!\!\!\!&&3\times \left\{3\times \left[2\times \left(+\frac{1}{6}\right)^{\!2}_{} - \left(-\frac{1}{3}\right)^{\!2}_{} - \left(+\frac{2}{3}\right)^{\!2}_{}\right] \times \left(+ \frac{1}{3}\right) \right.\nonumber\\
[2mm]
\!\!\!\!&&\left.+\left[2\times \left(-\frac{1}{2}\right)^2_{} - \left(-1\right)^2_{} \right]\times \left(-1\right) \right\}  =0\,.
\end{eqnarray}

The $U(1)_Y^{}-U(1)_{B-L}^{}-U(1)_{B-L}^{}$ anomaly is 
\begin{eqnarray}
&&3\times \left\{3\times \left[2\times \left(+\frac{1}{6}\right) -\left(-\frac{1}{3}\right)-\left(+\frac{2}{3}\right)\right]\times \left(+\frac{1}{3}\right)^{\!2}_{} \right.\nonumber\\
[2mm]
&&\left.+\left[2\times \left(-\frac{1}{2}\right) -\left(-1\right)\right]\times \left(-1\right) \right\}=0\,.
\end{eqnarray}

The $U(1)_{B-L}^{}-U(1)_{B-L}^{}-U(1)_{B-L}^{}$ anomaly is 
\begin{eqnarray}
&&3\times \left\{3\times \left[2\times \left(+\frac{1}{3}\right)^{\!3}_{}-\left(+\frac{1}{3}\right)^{\!3}_{}-\left(+\frac{1}{3}\right)^{\!3}_{}\right] \right.\nonumber\\
[2mm]
&&\left.+\left[2\times \left(-1\right)^{3}_{} -\left(-1\right)^{3}_{}\right] \right\} -2\times \left(-2\right)^{3}_{}  \nonumber\\
[2mm]
&&-\left(\frac{1+\sqrt{17}}{2}\right)^{\!3}_{} -\left(\frac{1-\sqrt{17}}{2}\right)^{\!3}_{}=0\,.
\end{eqnarray}

The graviton-graviton-$U(1)_{B-L}^{}$ anomaly is
\begin{eqnarray}
&&3\times \left\{3\times \left[2\times \left(+\frac{1}{3}\right)-\left(+\frac{1}{3}\right)-\left(+\frac{1}{3}\right)\right] \right.\nonumber\\
[2mm]
&&\left.+\left[2\times \left(-1\right) -\left(-1\right)\right] \right\} -2\times \left(-2\right) \nonumber\\
[2mm]
&&-\left(\frac{1+\sqrt{17}}{2}\right) -\left(\frac{1-\sqrt{17}}{2}\right)=0\,.
\end{eqnarray}

\section{The full scalar potential}

The full scalar potential of the renormalizable models is given by
\begin{eqnarray}
\label{potential}
V &=& \mu_\phi^2 \phi^\dagger_{}\phi+\lambda_\phi^{}\left(\phi^\dagger_{}\phi\right)^2_{} +\mu_\xi^2 \xi^\dagger_{}\xi + \lambda_\xi^{} \left(\xi^\dagger_{} \xi\right)^2_{} + \lambda_{\phi\xi}^{}\phi^\dagger_{}\phi \xi^\dagger_{}\xi\nonumber\\
[2mm]
&&+\sum_{a=1,...}^{}M_{\eta_a}^2 \eta^\dagger_{a}\eta_a^{} + \sum_{a,b,c,d=1,...}^{}\lambda_{abcd}^{} \eta_a^\dagger \eta_b^{}  \eta_c^\dagger \eta_d^{} \nonumber\\
[2mm]
&&+\sum_{a,b=1,...}^{}  \left(\lambda _{ab\phi}^{} \eta^\dagger_a \eta_b^{} \phi^\dagger_{}\phi + \lambda'^{} _{ab\phi} \eta^\dagger_a \phi^{} \phi^\dagger_{}\eta_b^{}\right.\nonumber\\
[2mm]
&&
\left.+\lambda_{ab\xi}^{} \eta^\dagger_{a}\eta_b^{} \xi^\dagger_{}\xi\right)+\sum_{a=1,...}^{} \rho_{a}^{}\left( \xi \eta_a^\dagger \phi +\textrm{H.c.}\right).
\end{eqnarray} 
For convenience and without loss of generality, we have chosen the mass matrices $M_\eta^2$ to be real and diagonal. In this basis, we have further rotated the parameters $\rho_a^{}$ to be real, i.e. $\rho_a^{}=\rho_a^\ast$.

To ensure the scalar potential bounded from below, the quartic couplings in Eq. (\ref{potential}) should fulfil the following conditions, 
\begin{eqnarray}
\label{bbelow}
&&\lambda_\phi^{}\,,~\lambda_\xi^{}\,,~\lambda_{aaaa}^{}>0\,,\nonumber\\
[2mm]
&& \lambda^{}_{aabb }+ \lambda^{}_{abba} +2\sqrt{\lambda_{aaaa}^{}\lambda_{bbbb}^{}}>0\,,\nonumber\\
[2mm]
&&\lambda_{\phi\xi}^{}+2\sqrt{\lambda_\phi^{}\lambda_\xi^{}} >0 \,,\nonumber\\
[2mm]
&& \lambda^{}_{aa \phi }+ \lambda'^{}_{aa \phi }+ 2\sqrt{\lambda_\phi^{}\lambda_{aaaa}^{}}>0\,,\nonumber\\
[2mm]
&& \lambda^{}_{aa \xi } +2\sqrt{\lambda_\xi^{}\lambda_{aaaa}^{}}>0 \,.
\end{eqnarray} 
Moreover, the perturbativity condition requires all quartic couplings less than $4\pi$, i.e.
\begin{eqnarray}
\label{pertur}
\lambda_\phi^{}\,,~\lambda_\xi^{}\,,~\lambda_{\phi\xi}^{}\,,~\lambda_{abcd}^{}\,,~ \lambda^{}_{ab \phi }\,,~ \lambda'^{}_{ab \phi }\,,~ \lambda^{}_{ab \xi } <4\pi \,.
\end{eqnarray}

On the other hand, the cubic couplings $\rho_a^{}$ will lead to corrections on these quartic couplings. Specifically, by integrating out the heavy Higgs doublets $\eta_a^{}$ at tree level, we can easily read
\begin{eqnarray}
\delta \lambda_{\phi\xi}^{} &=&-\sum_{a=1,...}^{}\frac{\rho_a^2 }{M_{\eta_a}^2}\,.
\end{eqnarray} 
At one-loop order, the corrections can be induced by some box diagrams, which are not shown for simplicity. We can roughly estimate
\begin{eqnarray} 
\delta \lambda_\phi^{} &\sim & \frac{1}{16\pi^2_{}}\sum_{a,b=1,...}^{}\frac{\rho_a^2 \rho_b^2}{M_{\eta_a}^2 M_{\eta_b}^2} \,,\\
\delta \lambda_\xi^{} &\sim &\frac{1}{16\pi^2_{}}\sum_{a,b=1,...}^{}\frac{\rho_a^2 \rho_b^2}{M_{\eta_a}^2 M_{\eta_b}^2}\,,\\
\delta \lambda_{\phi\xi}^{} &\sim &\frac{1}{16\pi^2_{}}\sum_{a,b=1,...}^{}\frac{\rho_a^2 \rho_b^2}{M_{\eta_a}^2 M_{\eta_b}^2}\,,\\
\delta \lambda_{abcd}^{} &\sim &\frac{1}{16\pi^2_{}}\sum_{a,b,c,d=1,...}^{}\frac{\rho_a^{} \rho_b^{} \rho_c^{} \rho_d^{} }{M_{\eta_a}^{} M_{\eta_b}^{} M_{\eta_c}^{} M_{\eta_d}^{}}\,.
\end{eqnarray} 
Then the conditions (\ref{bbelow}) and (\ref{pertur}) impose the following constraint on the cubic couplings $\rho_a^{}$, i.e.
\begin{eqnarray}
\sum_{a=1,...}^{}\frac{\rho_a^2 }{M_{\eta_a}^2} < 4\pi\,.
\end{eqnarray}

\end{document}